\documentclass[reprint,aip,rsi,amsmath,amssymb,nofootinbib]{revtex4-2}
\usepackage{amssymb}
\usepackage{lipsum}
\usepackage{amssymb}
\usepackage{bbold}
\usepackage{amssymb}
\usepackage{graphicx}
\usepackage{graphics}
\usepackage{amsmath}
\usepackage{placeins}
\usepackage{hyperref}
\usepackage[dvipsnames]{xcolor}
\usepackage{wasysym}
\usepackage{soul}
\usepackage{float}
\usepackage{array}
\usepackage{tabulary}
\usepackage{caption}
\usepackage{fancyhdr}
\usepackage{wrapfig}

\begin{document}
\title{Computationally Assessing Diamond as an Ultrafast Pulse Shaper for High Power Ultrawide Band Radar}

\author{Christopher S. Herrmann}
\affiliation{Department of Electrical and Computer Engineering, Michigan State University, 428 S. Shaw Ln., East
Lansing, MI 48824, USA}

\author{Joseph Croman}
\affiliation{High Power Microwave Section, Code 5745, Naval Research Laboratory, Washington, DC 20375, USA}

\author{Sergey V. Baryshev}
\email{serbar@msu.edu}
\affiliation{Department of Electrical and Computer Engineering, Michigan State University, 428 S. Shaw Ln., East
Lansing, MI 48824, USA}
\affiliation{Department of Chemical Engineering and Materials Science, Michigan State University, 428 S. Shaw Ln., East
Lansing, MI 48824, USA}


\begin{abstract}
Diamond holds promise to reshape ultrafast and high power electronics. One such solid-state device is the diode avalanche shaper (DAS), which functions as an ultrafast closing switch where closing is caused by the formation of the streamer traversing the diode much faster than 10$^7$ cm/s. One of the most prominent applications of DAS is in ultrawide band (UWB) radio/radar. Here we simulate a diamond-based DAS and compare the results to a silicon-based DAS. All DAS were simulated in mixed mode as ideal devices using the drift-diffusion model. The simulations show that diamond DAS promises to outperform Si DAS when sharpening kilovolt nanosecond input pulse. The breakdown field and streamer velocity ($\sim$10 times larger in diamond as compared to those in Si) are likely to be the major reasons enabling kV sub-50 ps switching using diamond DAS.
\end{abstract}

\maketitle

\section{Introduction}
\label{sec:introduction} 
Ultrawide band (UBW) radio/radar has become a promising area of radio frequency technology and continues to make inroads in communication, sensing and vision across consumer, commercial and military sectors. 
UWB radar has a much wider frequency spectrum of transmitted pulses when compared to conventional radar and radio signals: bandwidth of UWB is often cited as $>$25\% of the center frequency versus $<$1\% in conventional radars.\cite{UWBbook1}
A unique feature of UWB radio is its immunity to passive environmental interference as well as to interference in spatially and spectrum crowded conditions where multiple systems operate simultaneously. This makes it a critical technology for autonomous and connected systems. UWB radar performance, characterized by detection resolution and range, depends directly on both the duration and the power level of the pulse provided to the system. The shorter the pulse (the wider the band) the better the resolution. The higher the average power the longer the range.


Ultrashort pulses can be obtained using semiconductor step recovery diode (SRD) or Schottky diode,\cite{srd} but they can only switch a few volts and thus are low peak-power. In the 1980's, Si technology for opening (drift SRD, or DSRD) \cite{dsrd85} and closing switches (diode avalanche shaper, DAS) \cite{das81} was introduced. The DSRD and DAS are both $p$-$i$-$n$ diodes. When stacked, MW peak power, low jitter pulses have been demonstrated with leading and trailing edges in the sub-ns range.\cite{jitter} Advancing DSRD and DAS technology in terms of increasing the voltage rating and decreasing the switching time would be advantageous and critical to further push the envelope of UWB radar capabilities. For instance, recent amendments to spectrum management and telecommunication policies outline UWB radar operating in the 30 GHz band as essential for future vehicular radar technology, which requires temporal pulse length approaching 10 ps. Other important disciplines could greatly benefit, including ultrafast electro-optics and laser technologies,\cite{switch} discharge/plasma-assisted combustion and chemistry,\cite{Starik} bio-electro-magnetics,\cite{bioEM} and accelerator systems.\cite{Rukin} In the high-power application space, sufficiently mature DSRD and DAS devices may also be leveraged as high-tech UWB sources of intentional electromagnetic interference (IEMI).\cite{Giri}


Previous work on Si shapers\cite{Chudo} attempted to look beyond Si technology. 
Direct bandgap GaN and GaAs were ruled out. 
No reasonable semiconductor offers both improved breakdown field, high mobility and minority carrier lifetime except for diamond and SiC to potentially outperform Si. A comparison between Si, SiC and diamond is drawn in Table~\ref{matertable}. The most significant parameter is the ultimate differential voltage $dV/dt=E_{br}\times v_{s}$. Comparison between Refs.\onlinecite{JFOM, dsrd85} shows that the ultimate differential voltage is exactly the Johnson's figure of merit (JFOM). For a Si DSRD, the ultimate differential voltage is 2 V/ps. Ideal SiC and diamond DSRD opening switches can offer 90 and 200 V/ps, respectively. This simultaneously provides higher peak power in the GW range with a smaller count of diodes in the stack, a smaller footprint, and reduced cooling requirements. When paired with corresponding SiC and diamond DAS closing-switches, ultimate voltage rates between 900 to 2,000 V/ps can ideally be obtained (where a Si DSRD/DAS pair produces $\lesssim$20 V/ps). The team at Ioffe Institute has been actively exploring the potential of the SiC technology.\cite{SiCdsrd,SiCdas,grexJAP} Presented here is an early attempt to assess the potential of diamond for ultrafast high power switching and pulse shaping applications.

\begin{table*}[!]
 \centering
\caption{Comparison of Ultimate Material Properties}
\begin{tabular}{ l  c  c  c c}

                                                          & Diamond & Si & 4H-SiC\\
\hline
\hline
 Bandgap $E_g$ (eV)                                       & 5.47 & 1.1  & 3.2  \\
 Breakdown field $E_{br}$ (MV/cm)                         & 10   & 0.2 & 3    \\
 Electron Mobility $\mu_e$ (cm$^2$/V$\cdot$s)             & 4500 & 1450 & 900  \\
 Hole Mobility $\mu_h$ (cm$^2$/V$\cdot$s)                 & 3800 & 480  & 480  \\
 Electron Saturation Velocity $v_s$ ($\times$10$^7$ cm/s) & 2    & 1 & 3    \\
 Johnson's Figure of Merit $E_{br}\times v_s$ (V/ps)      & 200  & 2 & 90
\end{tabular}
\label{matertable}
\end{table*}    

\section{Pulse Shaping and Operating Principle of DAS}
\begin{figure*}[]
\begin{center}
\includegraphics[height=6.cm]{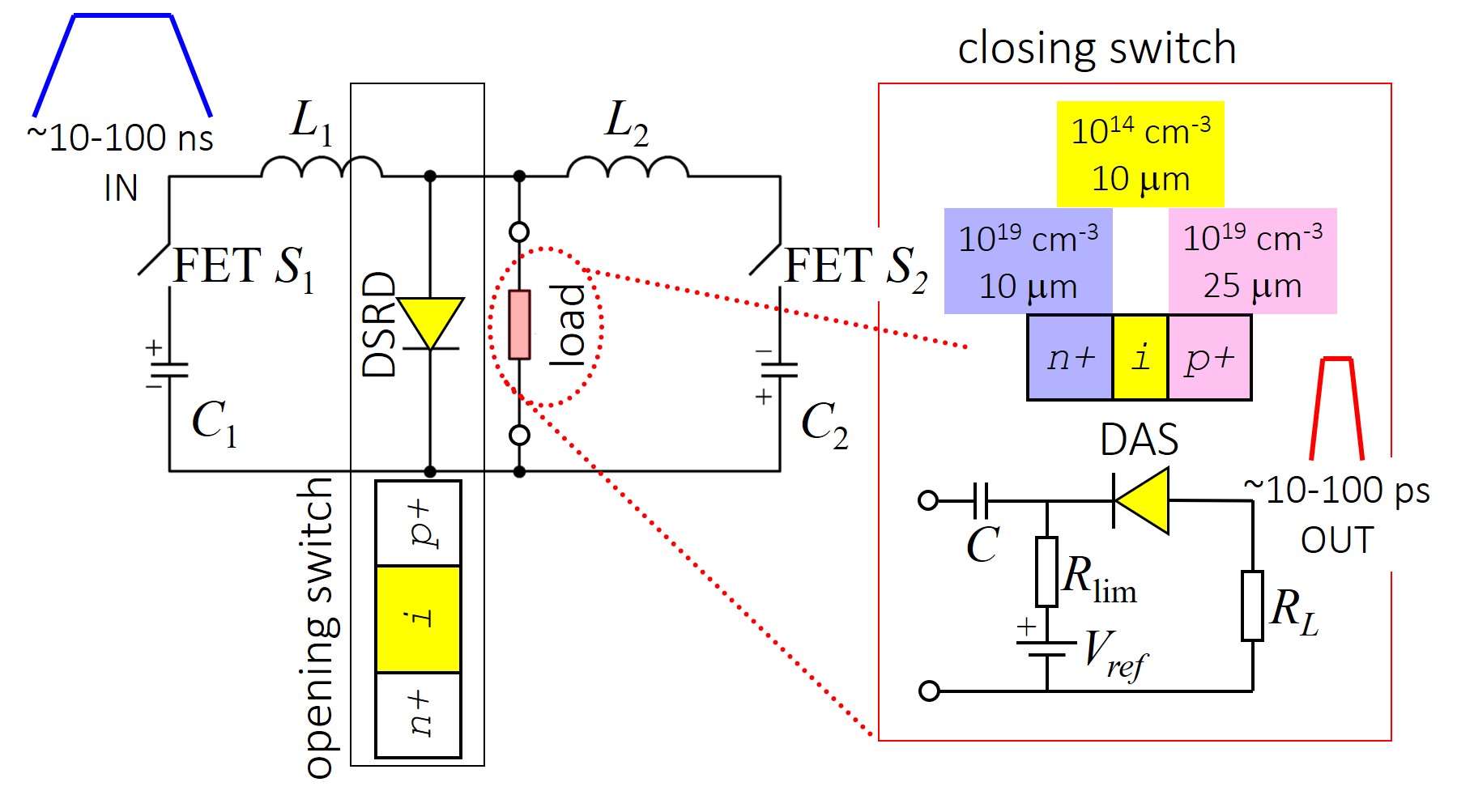}
\caption{A high-power ultrafast circuit based on closing and opening switches, adapted from Refs.\onlinecite{SiCdsrd, FociaRSI}. Simulated circuit is in the red box.}
\label{Circuit}
\end{center}
\end{figure*}

A basic inductive storage pulse shaping circuit arrangement is depicted in Fig.~\ref{Circuit}. The DSRD is an intermediate step between the high power field effect transistor (FET) that produces the high voltage nanosecond pulse for the DAS. The role of the DSRD is to properly trigger the DAS. The mating condition between the singular DSRD and DAS is the JFOM:\cite{grexJAP} in this context a DSRD and DAS made of the same semiconductor would give the best performance, but it is not a primary requirement. The DSRD shapes the pulse coming from the FET from roughly 10 ns to roughly 1 ns which is transferred to the DAS for final sub-ns shaping. Both the FET and the DSRD are transit time or JFOM limited. In the DAS, ultrafast power shaping to less than 1 ns is achieved through a plasma effect called a streamer, which traverses the diode much faster than a charge moving at saturated drift velocity $\sim$10$^7$ cm/s. The JFOM is now replaced by $E_{max}\times v_p$, where $E_{max}$ is the peak pulse (or  maximum) electric field of the “overvoltaged” diode such that $E_{max}$ is larger than the static $E_{br}$ and $v_p$ is the plasma streamer velocity that is larger than $v_s$. Modelling of the DSRD inside the inductive circuit with fast FET's is fairly straightforward and involves adjusting the depleted capacitance inside a standard PSpice diode model. In what follows, we do not consider or model a DSRD opening-switch and focus solely on the DAS closing-switch behavior. This is because analysis of a DAS made on a novel semiconductor such as diamond involves rich set of physics and circuit processes that must be solved self-consistently. We consider the DSRD as a hypothetical device that provides a triggering input pulse for the DAS at varied input $dV/dt$.

The pulse sharpening produced by the DAS is due to delayed ionization breakdown (in contrast to the common avalanche breakdown that takes place instantaneously) that forms a high charge plasma streamer travelling at a velocity much higher than the saturated drift velocity. In this case, the conventional transit time limitation is overcome. To form the streamer, fast overvoltaging (kilovolts on the nanosecond scale, reported as the voltage ramp rate $dV/dt$) of a reversely biased diode must be provided. With $dV/dt$$\gtrsim$1 V/ps, where the state-of-the-art Si DSRD technology can produce 2 V/ps,\cite{dsrd85,FociaMS} avalanche breakdown does not instantaneously occur despite an applied voltage that can reach a few times higher than the dc breakdown voltage. It should be noted, a stack of Si DSRDs can produce $dV/dt$ much higher than 2 V/ps. Important reference here is a top of the line generator consisting of stacked Si DSRDs that was able to deliver 80 kV in 1.5 ns, yielding 50 V/ps.\cite{80kV}

\begin{figure*}[!]
\begin{center}
\includegraphics[height=6.5cm]{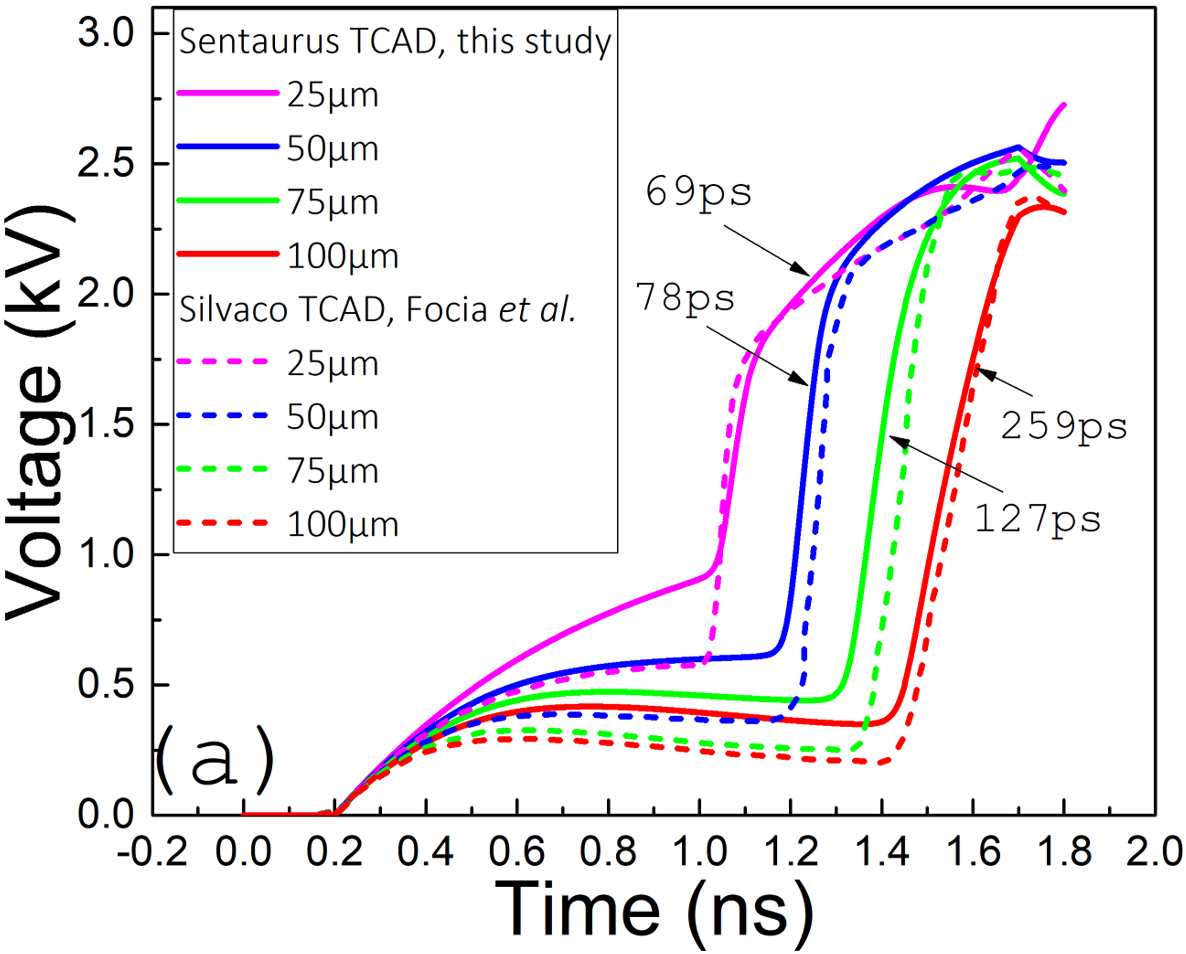}\includegraphics[height=6.5cm]{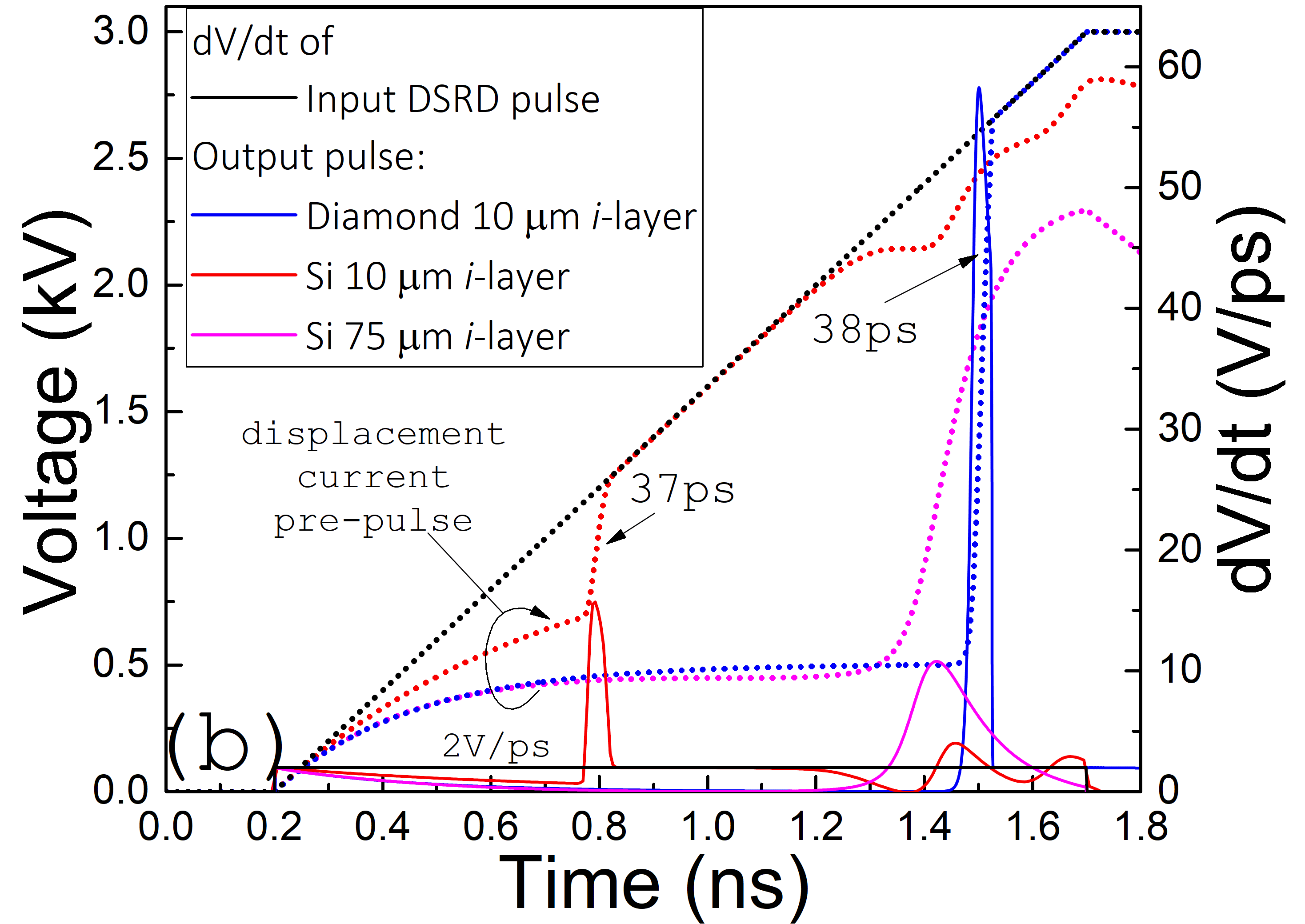}
\caption{Transient analysis of Si and diamond DAS: (a) benchmarking of the Sentaurus TCAD model to the previous results obtained for Si DAS;\cite{FociaMS} (b) comparison between thin base, 10 $\mu$m, Si and diamond DAS and thick base, 75 $\mu$m, Si DAS. Arrows denote FWHM of the output voltage rate $dV/dt$. Black, blue, red and magenta dotted lines represent input pulse from a DSRD, output pulse from the 10-$\mu$m base diamond DAS, output pulse from the 10-$\mu$m base Si DAS, and output pulse from the 75-$\mu$m base Si DAS, respectively. Black, blue, red and magenta solid lines are time derivatives of the corresponding color dotted lines.}
\label{DASvsSAS}
\end{center}
\end{figure*}

First, a displacement current $j_d$ associated with time varying electric field $\varepsilon\varepsilon_0\partial E/\partial t$, flows through the diode, forming a pre-pulse voltage across the load (see Fig.~\ref{DASvsSAS}). As an electric field is established across the DAS, charges accelerate and produce electron-hole plasma in the undoped/low doped, $i$--layer of the device. This plasma continues to accumulate as $n_p=n_0e^{v_s\alpha(E)t}$, as long as the device does not break down. The rate of this plasma generation is directly related to the field dependent ionization coefficient $\alpha(E)$ which is a material property. When the plasma streamer forms, current flows freely and at plasma velocity $v_{p}$ that is $\sim$10--100 times higher than the saturated drift velocity $v_s$. $v_p$ takes the form\cite{grexJAP}
\begin{equation}
v_p\sim v_s\frac{n_p}{n_d}\bigg(\frac{E_{max}}{E_{br}}-1\bigg)\frac{1}{ln(\frac{n_p}{n_0})},
\label{eq1}
\end{equation}
\noindent here $n_0$ ($\gtrsim 10^9$ cm$^{-3}$) is the initial background charge concentration in the depleted base (same as $i$-layer), $n_d$ (10$^{14}$ cm$^{-3}$ in our case) is the doping concentration in the base layer, $E_{max}$ is the maximum electric field of the overvoltaged $p$-$i$-$n$ diode such that $E_{max}/E_{br}$ is between 1 and 2. $n_p$ is the plasma charge concentration that can be found as $n_p\sim\frac{\alpha\varepsilon\varepsilon_0 E_{br}}{q}$, where $q$ is elementary charge, $\varepsilon\varepsilon_0$ is material permittivity. For diamond, it results in $v_p\gtrsim 10^8$ cm/s. Streamers causing lightning are known to move as fast as $\sim$10$^7$-10$^8$ cm/s.\cite{lightning} Major factors that affect switching performance include thickness of the diode base layer $L_i$, breakdown voltage $V_{br}$, as well as the input voltage ramp rate $dV/dt$, a limitation that comes from the input pulse generator:

\begin{equation}
t_{lim1}=\frac{L_{i}}{v_{p}},
\label{eq2}
\end{equation}

\begin{equation}
t_{lim2}=\frac{V_{br}-V_{ref}}{dV/dt},
\label{eq3}
\end{equation}
where $V_{ref}$ is a dc pre-bias. For $L_i$=10-100 $\mu$m, switching takes place on a sub-nanosecond time scale. This is the only non-optical method allowing for conductivity modulation of a semiconductor structure at 100 ps or faster.

\section{Modelling Methodology}
\label{sec:modelling} 
The main advantage of diamond over Si is two-fold. One is the streamer velocity: it can be estimated that for diamond its value is near or above 10$^8$ cm/s while for Si it is just above $v_s$ (3$\times$10$^7$ cm/s). Second, is that for a given thickness the breakdown voltage of the diamond DAS is larger, and 
diamond $\alpha(E)$ dependence is shifted to higher fields, resulting in faster multiplication and thus a thinner base in the $p$-$i$-$n$ structure. Overall, this means that switching enabled by a diamond-based DAS can provide for a much higher output voltage rate as compared to Si DAS. Evidence for this is seen in Fig.~\ref{DASvsSAS}. The base $i$-layer is intentionally made ten/twenty-fold thinner than usually anticipated,\cite{das81,FociaIEEE} and results are compared to an optimized Si DAS design produced by work at the University of New Mexico by Focia \textit{et al.}\cite{FociaMS} Replicating device doping and geometry studied by Focia \textit{et al.} allowed for the cross-checking of our computational approach for the Si DAS (Fig.~\ref{DASvsSAS}a). All results presented in Fig.2 were obtained using an input pulse with a 3 kV peak voltage and  1.5 ns linear rise time generated from a hypothetical DSRD, and was identical for Si and diamond DAS devices.

After validating our computational approach, diamond was implemented as a custom material based on Watanabe's $\alpha(E)$ data.\cite{Watanabe} Watanabe's data describes diamond with $E_{br}\sim10^6$ V/cm. Since the considered diamond DAS is a heavily punched-through $p$-$i$-$n$ diode, its static breakdown voltage $U_{br}$ can be estimated as the product of $E_{br}$=10$^6$ V/cm and $i$-layer thickness of 10 $\mu$m, i.e. gives 1 kV. Therefore, the incoming 3 kV pulse provides for 3-fold overvoltaging ($E_{max}/E_{br}\sim 3$) and guarantees the formation of ultrafast impact ionisation plasma front. Early comparisons seen in  Fig.~\ref{DASvsSAS}b favors the 10 $\mu$m diamond device over 75 $\mu$m Si device:\cite{FociaIEEE} peak output voltage rate (right axis) of the diamond DAS is six times that of the Si DAS and the diamond DAS exhibits a higher hold off voltage. Each dotted trace represents a transient voltage curve with the time derivatives of these curves being represented by the solid lines. The black curve shows the input pulse applied to the DAS while the colored curves represents devices simulated. Comparing results of Fig.~\ref{DASvsSAS}a and b, it is clear that thinner base provides for better sharpening but at the cost of smaller output voltage: reducing the $i$ layer thickness of the Si DAS by 10 times improved switching from 259 to 37 ps but output voltage reduced from 2 to 0.5 kV. This is while the diamond DAS with a 10 $\mu$m $i$ layer switched in 38 ps providing for 2.1 kV of output voltage.

The drift-diffusion approach was used to model the physical behavior of the DAS in three dimensions using the Synopsys Sentaurus TCAD software. This 3D model was configured to simulate constant carrier mobility, carrier velocity saturated at high fields, and avalanche dynamics in accord with the Van Overstraet de Man model.\cite{Vandeman} Tunneling-assisted impact ionization was not simulated as it is expected play no role for a DAS based on a wide bandgap material.\cite{grexJAP} Thermodynamic equations and device self-heating were not modeled and all simulations took place at 300 K. The DAS devices were simulated in the mixed mode which allowed for mixed physics and circuit SPICE-like transient analyses. The virtual circuit, shown in Fig.~\ref{Circuit}, consisted of a voltage source connected to the cathode of the diode with the anode connected to a 50 $\Omega$ load resistor. The amplitude of the hypothetical DSRD-based voltage source ramped at a constant, linear rate, providing the input voltage with a rise rate of 1-100 V/ps for the DAS to sharpen. A dc voltage source and current limiting resistor were also applied to the cathode of the diode. This allows for the breakdown point to be tuned, see Eq.~\ref{eq3}; $V_{ref}$ was kept constant at 10 V. 

\section{Results}

To assess the advanced pulse sharpening capability enabled by diamond, the sharpened output voltage rates $dV/dt$ are compared as a performance figure. There is of course, no singular metric that can ultimately capture the overall performance of a DAS as different applications require certain characteristics to be a priority. Still, the time-derivative of the output voltage curves gives an impression of performance improvement between Si and diamond devices as seen in Fig.~\ref{DASvsSAS}. Comparison between 10 $\mu$m and 75 $\mu$m Si devices show no significant improvement in terms of peak output $dV/dt$. With all other parameters held equal, the thinner base Si DAS provides for faster switching, consistent with Eq.~\ref{eq2}, but hold off voltage is reduced; and vice versa for the thicker base DAS. To further explore $dV/dt$ performance, we chose to compare a 10 $\mu$m base DAS made of Si and diamond (shown in Fig.~\ref{Circuit}). The Si DAS has light $n$-doping in the $i$-layer, like in the classical case.\cite{das81} The diamond DAS has light $p$-doping to reflect upon intentional or unintentional doping with boron that has lowest activation energy among all known dopants. The cross-sectional area of DAS was 1 mm$^2$.

\begin{figure}[h]
\begin{center}
\includegraphics[height=7.5cm]{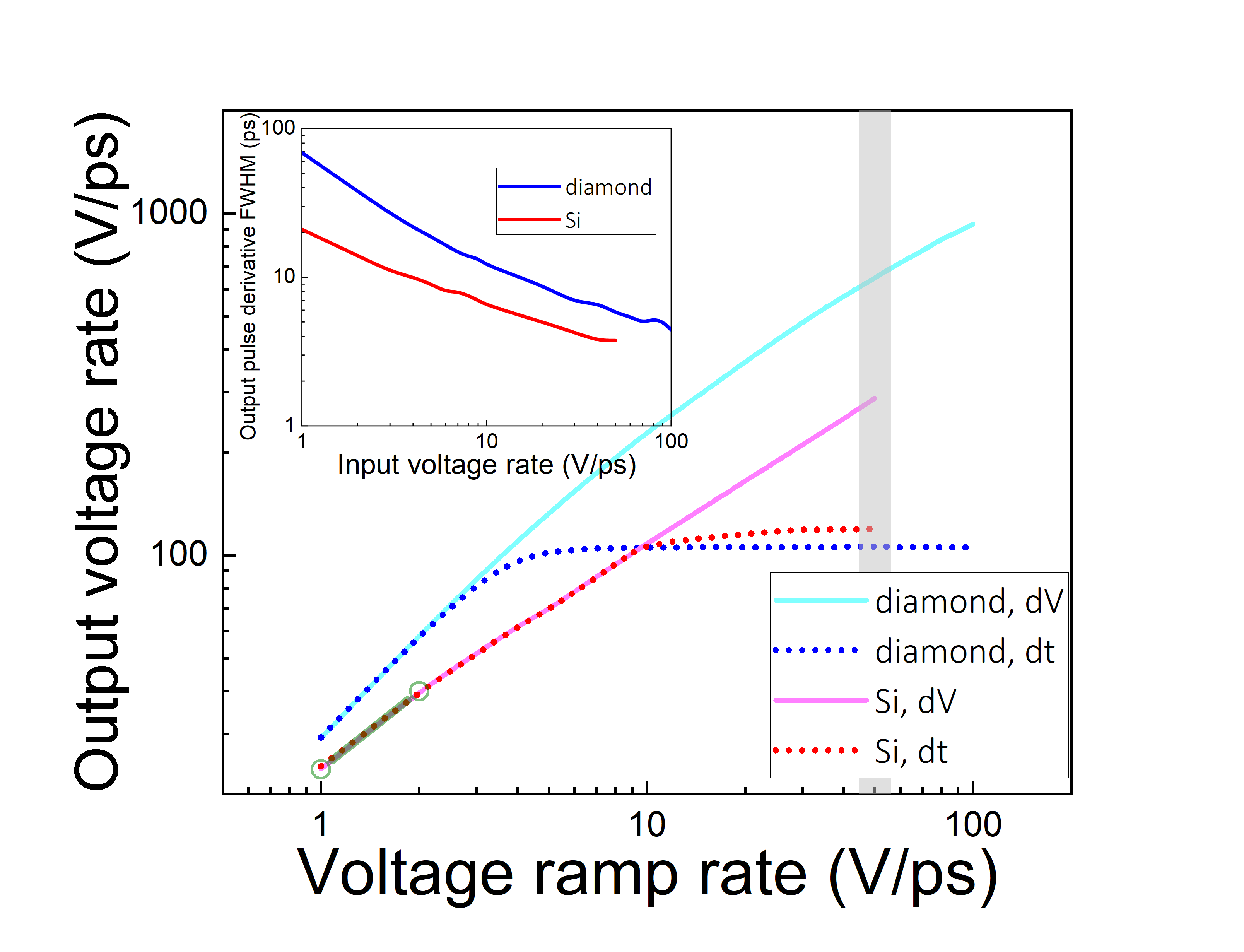}
\caption{Computed peak output voltage rate of DAS versus input voltage ramp rate from a hypothetical diamond DSRD producing up to 100 V/ps. Performance difference of changing peak amplitude $dV$ or pulse duration $dt$ is emphasized. The difference between single mated Si vs diamond DSRD/DAS pair is highlighted: the green dumbbell curve should be compared to the full range solid cyan curve. The inset compares resultant switching time (FWHM of the output voltage rate $dV/dt$) as a function of input voltage ramp rate for Si and diamond DAS.}
\label{Fig:PeakOutputVsInput}
\end{center}
\end{figure}

The performance of the device is dependent on the voltage ramp rate, $dV/dt$, of the input pulse seen in Fig~\ref{Fig:PeakOutputVsInput}. The slope of the black line shown in Fig.~\ref{DASvsSAS}b represents this value and is characterized by the ultimate pulse magnitude, $dV$, and the pulse duration, $dt$. The value $dV/dt$ can be changed by varying $dV$ and holding $dt$ constant, or vice versa. These methods are not equivalent. The solid lines in Fig.~\ref{Fig:PeakOutputVsInput} show simulations that were conducted with a constant $dt$ and increasing $dV$. There is a clear linear relationship between the $dV/dt$ of the input pulse and the peak output voltage rate. The dotted lines in Fig.~\ref{Fig:PeakOutputVsInput} show simulations that were conducted with shortening $dt$ and constant $dV$ values. Initially, the two methods remain consistent until diverging and leveling off. This divergence is thought to be due to the crystal lattice, which is to be ionized, being unable to adequately respond to the sub-ns pulses. The crystal lattice does not have enough time to react to the many-kV pulse applied to it (phononic phenomena are on $\sim$ns time scale). This divergence occurs despite the device being adequately reverse biased in accord with the overvoltaging approximation seen in Section \ref{sec:modelling}. The presentation of Fig.~\ref{Fig:PeakOutputVsInput} indicates that future input pulse generators should prioritize ultimate voltage magnitude, but short input pulse duration is still required. A hypothetical diamond DSRD would excel at this.

Fig.~\ref{Fig:PeakOutputVsInput} shows that for every given input voltage ramp rate, diamond exhibited a higher output voltage rate. Ultimate input voltage ramp rates ranged from 1 to 100 V/ps, reflecting upon near future advances in diamond DSRD technology that could enable input pulses with such characteristics, see JFOM in Table~\ref{matertable}. Since Si based DSRD technology can ultimately produce not more than 2 V/ps,\cite{dsrd85} multiple Si DSRDs must be stacked together to overcome this 2 V/ps limitation. Performance mismatches of stacked devices may quickly diminish the overall performance. Recent review by Rukin\cite{Rukin} still cites the 80 kV 1.5 ns DSRD generator designed at Ioffe Institute\cite{80kV} as the top of the line. From this, 50 V/ps is used as a cut-off input $dV/dt$ that can be produced by the stacked Si technology. Tying back to the benefits of diamond, an ideal singular diamond DSRD could offer 200 V/ps allowing for a smaller count of diodes in the stack while promising greater relief on cooling requirements. Therefore, it is instructional to compare the performance of a pair of a single Si DSRD mated with a single Si DAS (semitransparent dumbbell line in Fig.~\ref{Fig:PeakOutputVsInput}) with that made of diamond (full range solid cyan curve in Fig.~\ref{Fig:PeakOutputVsInput}). One can see that the simulations predict orders of magnitude improved performance of diamond pulsed power device generation, where total count of a few diamond diodes can produce the effect that requires several tens or even hundreds DSRD or DAS made of silicon.

On a $log$-$log$ scale, Fig.~\ref{Fig:PeakOutputVsInput} demonstrates that for every given input voltage ramp rate the output voltage rate increased 30 to 10 times (consistent with the shaping ratio $\frac{E_{max}\times v_p}{E_{br}\times v_s}$), providing an ultimate 1 kV/ps at the output when receiving an input of 0.1 kV/ps from a hypothetical ideal diamond DSRD. Given our simulation settings, such ultimate switching takes place on a 5 ps time scale (inset in Fig.~\ref{Fig:PeakOutputVsInput}) which would greatly expand the performance of a UWB radar. Such sharpening is enabled by operating the $p$-$i$-$n$ structure with a thin base layer, which comes at the expense of the delivered peak voltage that dropped from 150 kV to 5 kV, still producing half a MW of peak power. As mentioned earlier, designing a DAS has to start with application requirements with primary consideration being given to the trade-off between high peak voltage and switching time, the trade-off clearly seen in Fig.~\ref{DASvsSAS}a,b. Additional parameters of optimization also include varied doping levels and cross section area.

\section{Conclusion and Outlook}
It was shown computationally, that a DAS constructed from diamond exhibits superior performance when compared to identical diode structure made of Si: the output voltage rate was more than 3 times higher than that of the Si device which results in 10-fold increase in the output peak power for the same switching time. Special attention was paid to thin base devices to attest them for ultrafast high peak power pulse shapers in order to push the envelope in UWB radar applications. This demonstrates that 5-10 ps MW scale pulses are feasible with single crystal diamond homoepitaxy and efficient $n$-doping strategies. Diamond homoepitaxy has been continuously improving, however problems with $n$-type doping persist with many dopants like nitrogen, phosphorus, sulfur due to activation energy of these dopants being much higher than 26 meV (thermal energy at room temperature). Technologically, a canonical vertical $p$-$i$-$n$ diamond DAS and DSRD $n^+$ layers can be obtained through fabricating so-called merged diode,\cite{merge} depositing bandgap matched $n$-type AlGaN alloy or high conductivity $n$-type nano-crystalline diamond (NCD).\cite{Tanvi} The last NCD-based strategy was successfully implemented to fabricate high quality rectifying $p$-$n$ junctions were $n$-type NCD was deposited on $p$-type single crystal diamond,\cite{NCD-SCD} SiC\cite{NCD-SiC} and even Si.\cite{NCD-Si}

\bibliography{das}

\end{document}